\documentstyle[12pt]{article}
\let\Large=\large
\let\large=\normalsize
\begin{document}
\begin{flushright}
CAMS/99-01\\
hep-th/9903046   
\end{flushright}
\vspace{.5cm}
\begin{center}
\baselineskip=16pt {\Large {\bf Calabi-Yau Black Holes and Enhancement \\
of Supersymmetry in Five Dimensions}}
\vskip 1 cm 
{A. H. Chamseddine and \,W. A. Sabra }\footnote{
e-mail: chams@aub.edu.lb, {ws00@aub.edu.lb}} \vskip 1cm \centerline{{\em
Center for Advanced Mathematical Sciences (CAMS),} } \centerline{{\em and}} 
\centerline{{\em Physics Department, American University of Beirut, Lebanon}}
\end{center}

\vskip 1 cm \centerline{\bf ABSTRACT} \vspace{-0.3cm}

\begin{quote}
BPS electric and magnetic black hole solutions which break half of
supersymmetry in the theory of $N=2$ five-dimensional supergravity are
discussed. For models which arise as compactifications of M-theory on a
Calabi-Yau manifold, these solutions correspond, respectively, to the two
and five branes wrapping around the homology cycles of the Calabi-Yau
compact space. The electric solutions are reviewed and the magnetic
solutions are constructed. The near-horizon physics of these solutions is
examined and in particular the phenomenon of the enhancement of
supersymmetry. The solutions for the supersymmetric Killing spinor of the
near horizon geometry, identified as $AdS_{3} \times S^{2}$ and $AdS_{2}
\times S^{3}$ are also given. \vspace{2truecm}
\end{quote}

\newpage

\section{Introduction}

Considerable progress has been made in the study of BPS black hole states
of the low-energy effective actions of compactified string and $M$-theory 
\footnote{for reviews see for example \cite{review}}. This was mainly motivated by the
important role that these states play in the understanding of the
non-perturbative structure behind string theory. Toroidal compactifications
of string theory give rise to vacua with $N=4$ and $N=8$ supersymmetry and
BPS states of these models are restricted by the large supersymmetry.
Corrections to these solutions and their entropies can only
arise from higher loops, as the lowest order corrections are
known to vanish. Four and five dimensional $N=2$ supergravity models coupled
to vector and hyper-multiplets can arise, for example, from type II string
and M-theory compactified on a Calabi-Yau threefold. An important step in
the study of $N=2$ black holes was the realisation that their entropy is
given in terms of the extremum of the central charge and that the scalar
fields of the theory take fixed values at the horizon independent of their
boundary values at spatial infinity \cite{feka}.

The perturbative and nonperturbative corrections for the $N=2$ supergravity
models seem to make the study of their BPS black hole solutions more intricate.
However, the rich geometric structure of these theories \cite{special, very}
considerably simplifies the analysis of their black hole solutions. For
instance, the black hole metric in four dimensions can be expressed in terms
of symplectic invariant quantities where the symplectic sections satisfy
algebraic constraints involving a set of constrained harmonic functions \cite
{three}. In a similar way to Einstein-Maxwell theory \cite{hh}, different
types of solutions can be obtained depending on the choice of harmonic
functions. The solutions also depend on the choice of the prepotential which
defines the theory. In five dimensions, the black hole solutions \cite{scs}
can be expressed in terms of the rescaled cubic homogeneous prepotential
which defines very special geometry \cite{very}.

The BPS solutions can be regarded as solitons interpolating between two
vacua: Minkowski flat space at infinity and $AdS_{p+2} \times S^{d-p-2}$ near the
horizon \cite{g, kp,gt}, where $p$ is the dimensionality of the solution.
Anti-de Sitter spaces and spheres are known to admit Killing spinors which
we refer to as geometric Killing spinors. At a generic point in space-time, the BPS solution breaks half of
supersymmetry. However, if at the near
horizon the supersymmetric Killing spinor, defined by the zero mode of the
gravitinio supersymmetry transformation, can be identified with the geometric
spinor, then supersymmetry is enhanced. This restoration of supersymmetry
have been discussed in the literature for the cases where the near horizon
geometry was identified with the spaces $AdS_{2} \times S^{2}$ \cite{g,kp}, 
$AdS_{5} \times S^{5}$, $AdS_{3} \times S^{7}$ , and $AdS_{7} \times S^{3}$ 
\cite{gt}. In five dimensional supergravity models with vector multiplets,
our concern in this paper, the restoration of supersymmetry of a
double-extreme black hole solutions \cite{double} near the horizon in four and five
dimensions has been discussed in \cite{feka, cfgk} and also for the rotating
case in \cite{raja}. The analysis of the pure five dimensional supergravity case (without
vector multiplets) was performed in \cite{gkltt}.

General BPS black hole solutions which break half of supersymmetry of $d=5,$ 
$N=2$ supergravity theory with an arbitrary number of vector supermultiplets,
were obtained in \cite{scs}. There it was also demonstrated that the entropy
of these black holes is given in terms of the extremised central charge. The  
near horizon geometry of these solutions is the same as that of the double-extreme
solutions in accordance with the attractor behaviour. The main result of
this paper is the analysis of general magnetic-string solutions\footnote{ a magnetic string solution in five dimensions was obtained as a compactification of an intersection of 3 M-5-branes and a boost along the common string in \cite {klaus}}, which in
compactified M-theory on a Calabi-Yau manifold, correspond to five branes
wrapped around the four cycles of the Calabi-Yau compact space. The
enhancement of supersymmetry at the near horizon of these solutions is shown
and we derive the expression of the Killing spinor of the maximally
supersymmetric horizon. The analysis of \cite{cfgk} is also completed by
obtaining the form of the Killing spinor for the near horizon geometry for
the electric BPS solution given by $AdS_2\times S_3$.

The bosonic action of five dimensional $N=2$ supergravity coupled to vector
multiplets is given by 
\begin{equation}
e^{-1} {{\cal L} }= -{\frac{1}{2}} R - {\frac{1}{4}} G_{IJ} F_{\mu\nu} {}^I
F^{\mu\nu J}-{\frac{1}{2}} g_{ij} \partial_{\mu} \phi^i \partial^\mu \phi^j +
{\frac{e^{-1}}{48}} \epsilon^{\mu\nu\rho\sigma\lambda} C_{IJK}
F_{\mu\nu}^IF_{\rho\sigma}^JA_\lambda^K.
\label{action}
\end{equation}
The theory of five dimensional $N=2$ supergravity with vector
supermultiplets was considered in \cite{gst}. The compactification of $N=1$, $D=11$ supergravity \cite{original},
i. e., the low energy limit of M-theory, down to five dimensions on Calabi-Yau
3-folds ($CY_{3}$), results in a $d=5, N=2$
supergravity coupled to abelian vector supermultiplets based on the
structure of very special geometry \cite{very}. The five dimensional theory
contains the gravity multiplet, $h_{(1,1)}-1$ vector multiplets and 
$h_{(2,1)}+1$ hypermultiplets, ($h_{(1,1)},h_{(2,1)}$) are the Hodge numbers
of $CY_{3}$. The $(h_{(1,1)}-1)$-dimensional space of scalar components of
the abelian vector supermultiplets coupled to supergravity can be regarded
as a hypersurface of a $h_{(1,1)}$-dimensional manifold whose coordinates 
$X^I(\phi )$ are in correspondence with the vector bosons (including the
graviphoton). In five dimensions, an $N=2$ vector supermultiplet has a
single scalar field and therefore the scalar manifold is real. The defining
equation of the hypersurface is 
\begin{equation}
{\cal {V}}(X)=1
\end{equation}
and the prepotential ${\cal {V}}$ is a homogeneous cubic polynomial in the
coordinates $X^I(\phi )$ 
\begin{equation}
{\cal {V}}(X)=\frac{1}{6}\ C_{IJK}X^{I}X^{J}X^{K}=X^{I}X_{I}=1,\qquad
I, J, K=1,\ldots, h_{(1,1)}
\end{equation}
where $C_{IJK}$ are the topological intersection numbers and $X_I$ are the \lq\lq dual'' special coordinates (the four cycles).

The corresponding vector and scalar metrics in (\ref{action}) are completely encoded in the
function ${\cal V}(X)$ 
\begin{eqnarray}
G_{IJ} &=&-\frac{1}{2}\frac{\partial }{\partial X^{I}}\frac{\partial }{
\partial X^{J}}\ln {\cal {V}}(X)_{{|{\cal V}}=1},  \nonumber \\
g_{ij} &=&G_{IJ}\frac{\partial }{\partial \phi ^{i}}X^{I}\frac{
\partial }{\partial \phi ^{j}}X^{J}_{|{\cal V}=1}\  .  
\label{metric}
\end{eqnarray}

\section{Electric BPS Black Holes}
In the compactified M-theory on a Calabi-Yau manifold, electrically point
and magnetically charged string-like BPS states correspond to the two-
and five-branes of M-theory wrapped around the two- and four-cycles of the
Calabi-Yau space. Though the details of the low-energy Lagrangian depend
very much on the compactified Calabi-Yau space, the analysis of the BPS
solutions can be considerably simplified by the rich geometric structure
based on ``very special geometry'' \cite{very} underlying the $N=2$
five-dimensional theories with vector supermultiplets. Electrically charged
BPS solutions of the five dimensional theory have been analyzed in \cite{scs}
and will be reviewed in this section.

The supersymmetry transformation laws for the fermi fields in a bosonic
background are given by \cite{scs, gst} 
\begin{eqnarray}
\delta\psi_\mu &=& \Big({\cal {D}}_\mu+ {\frac{ i}{8}} X_I \Bigl(\Gamma_\mu{}
^{\nu\rho} - 4 \delta_\mu{}^ \nu \Gamma^\rho\Bigr) F_{\nu\rho}{}^I\Big)
\epsilon,  \nonumber \\
\delta \lambda _i &=&\Big({\frac{3}{8}}\partial_iX_I\Gamma^{\mu\nu}F_{\mu%
\nu}^I - {\frac{i}{2}} g_{ij} \Gamma^\mu \partial_\mu\phi^j \Big)\epsilon,
\label{stl}
\end{eqnarray}
where $\epsilon$ is the supersymmetry parameter and ${\cal D}_\mu$ the
covariant derivative.\footnote{
we use the metric $\eta^{ab}=(-,+,+,+,+)$, $\{{\Gamma^a, \Gamma^b}\}=2\eta^{ab}$, ${\cal {D}}_\mu=\partial_\mu+{\ \frac{1}{4}}
\omega_{\mu ab} \Gamma^{ab}$, $\omega_{\mu ab}$ is the spin connection, and $%
\Gamma^{{\nu}}$ are Dirac matrices and $\Gamma^{{a}_1{a} _2\cdots{a_n}}= {%
\frac{1}{n!}}\Gamma^{[{a_1}}\Gamma^{{a_2}} \cdots \Gamma^{{a_n}]}$.} The
metric for the BPS electric black hole solutions can be brought to the form 
\cite{scs} 
\begin{equation}
ds^2 =-e^{-4 U}(dt+w_mdx^m)^2 +e^{2 U} (d\vec{x})^2.
\label{magic}
\end{equation}
The vanishing of the supersymmetry transformations of the gravitino and gauginos, for the choice $\Gamma^0\epsilon=-i\epsilon$, imposes the following
relations for the supersymmetry variation parameter, $U$, $w_m$, the scalar fields and the gauge fields
\begin{eqnarray}
\partial_t\epsilon&=& 0,  \nonumber \\
(\partial_m+\partial_mU)\epsilon&=&0,  \nonumber \\
(X_IF^I_{mn})^-&=&(\partial_mQ_n-\partial_nQ_m)^-,  \nonumber \\
(X_IF^I_{tm})&=&-\partial_me^{-2U},  \nonumber \\
2(\partial_mUw_n-\partial_nUw_m) -{\frac{1}{2}}(\partial_mw_n-
\partial_nw_m)+ e^{2U}(X_IF_{mq}^I)\nonumber \\
-{\frac{1}{2}} e^{2U}(X_IF_{mq}^I)^*-(\partial_mUw_n-\partial_nUw_m)^*&=&0, 
\nonumber \\
{\frac{3}{2}}e^{-U}\partial_mX_I\partial_iX^I-G_{IJ}e^U
\partial_iX^IF_{tm}^J&=&0,  \nonumber \\
\Big(G_{IJ}\partial_iX^IF_{mn}^J+ {\frac{3}{2}}e^{-2U}(\partial_mX_Iw_n-
\partial_nX_I w_m)\Big)^-&=&0.
\end{eqnarray}
where $Q_n\equiv e^{-2U} w_n$ and $F^-_{mn}=F_{mn} - ^{*}F_{mn}.$ The above
conditions are satisfied for 
\begin{eqnarray}
F^I_{mn} &=&\partial_m(X^IQ_n)-\partial_n(X^IQ_m),  \nonumber \\
F^I_{tm}&=&-\partial_m(e^{-2U}X^I),\nonumber\\ 
(\partial_mw_n-\partial_nw_m)^{-}&=&0\nonumber\\
\epsilon &=& e^{-U}\epsilon_0.
\label{gf}
\end{eqnarray}
where $\epsilon_0$ is a constant spinor satisfying $\Gamma^0\epsilon=-i\epsilon$.

The various physical variables can be fixed in terms of space-time
functions, by solving the equations of motion for the gauge fields
\footnote{
notice that the Bianchi identities are trivially satisfied}. From the
Lagrangian (\ref{action}), we derive the following equation of motion for
the gauge fields, 
\begin{equation}
\partial_\nu(eG_{IJ}g^{\mu\rho}g^{\nu\sigma} F_{\rho\sigma}^J)={\frac{1}{16}}
C_{IJK}\epsilon^{\mu\nu\rho\sigma k} F_{\nu\rho}^JF_{\sigma k}^K.
\label{gaugeeqn}
\end{equation}
Using (\ref{gf}) and after some manipulations one obtains the solution 
\begin{equation}
e^{2U}X_I={\frac{1}{3}}H_I, \label{jennie}
\end{equation}
where $H_I$ are harmonic functions, $H_I=h_I+{\frac{q_I}{r^2}}$, $h_I$ are
constants and $q_I$ are electric charges.

It is convenient to express the solution in terms of the geometry of the
internal space, i. e., in terms of the cubic polynomial ${\cal V}$. Define
the rescaled coordinates 
\begin{equation}
Y_I=e^{2U}X_I, \qquad Y^I=e^{U}X^I,
\end{equation}
then the underlying very special geometry implies that 
\begin{equation}
e^{3U}={\cal V}(Y)={\frac{1}{3}}C_{IJK}Y^IY^JY^K,
\end{equation}
and thus the electric black hole solution metric has the form 
\begin{eqnarray}
ds^2&=&-{\cal V}^{-4/3}(Y)(dt+\omega_mdx^m)^2+ {\cal V}^{2/3}(Y)(d\vec{x})^2,
\nonumber \\
Y_I&=&{1\over 3}H_I,  \nonumber \\
F^I_{mn} &=&\partial_m({\cal V}^{-1}Y^Iw_n)-\partial_n({\cal V}^{-1}Y^Iw_m),
\nonumber \\
F^I_{tm}&=&-\partial_m({\cal V}^{-1}Y^I).  \label{ramzi}
\end{eqnarray}

Black holes with rotational symmetry can be constructed using the above
general solution. Changing to polar coordinates defined by, $x^{1}+ix^{2}=r\sin \theta e^{i\phi },$ 
$x^{3}+ix^{4}=r\cos \theta e^{i\psi }$, and specializing to solutions with rotational symmetry in the two orthogonal
planes, i. e., $w_{\phi }=w_{\phi }(r,\theta ),$ $w_{\psi }=w_{\psi
}(r,\theta ),w_{r}=\omega _{\theta }=0,$ then the self-duality condition of
the field strength of $w_{m}$ gives for a decaying solution, 
$w_{\phi }=-{\frac{\alpha }{r^{2}}}\sin ^{2}\theta ,$ $w_{\psi }={\frac{\alpha }{r^{2}}}
\cos ^{2}\theta .$ Therefore, the general form of the metric solution with rotational
symmetry in the two orthogonal planes is given by 
\begin{equation}
ds^{2}=-{\cal V}^{-4/3}(Y)\Big(dt-{\frac{\alpha \sin ^{2}\theta }{r^{2}}}d\phi +{\frac{
\alpha \cos ^{2}\theta }{r^{2}}}d\psi\Big)^{2}+{\cal V}^{2/3}\Big(dr^{2}+r^{2}d\Omega_3^{2})
\end{equation}
where 
\begin{equation}
d\Omega_3^{2}=(d\theta ^{2}+\sin ^{2}\theta d\phi ^{2}+\cos ^{2}\theta d\psi
^{2}).
\end{equation}

It was demonstrated in \cite{cfgk} that for double-extreme black
holes in five dimensions supersymmetry is enhanced at the near horizon where
the metric takes the form 
\begin{equation}
ds^{2}=-({\frac{2\hat{r}}{r_{0}}})^{2}dt^{2}+({\frac{2\hat{r}}{r_{0}}})^{-2}d%
{\hat{r}}^{2}+r_{0}^{2}d\Omega _{3}^{2},  \label{berlin}
\end{equation}
with the geometry $AdS_{2}\times S^{3}$. Also these results were extended to
the rotating case where similar conclusions hold \cite{raja}. In reality one can
obtain the stronger result that for a general electric BPS black hole in
five dimensions with vector supermultiplets, supersymmetry is restored at
the near-horizon. This is because both extreme and double-extreme black
holes have the same form of metric at the near horizon. This can be seen by
examining the general solution near the horizon, ($r\rightarrow 0$). There $
{\cal V}(Y)$ can be approximated as follows. 
\begin{equation}
{\cal V}_{hor}(Y)={\frac{1}{3}}(Y^{I}H_{I})_{hor}={\frac{1}{3}}(Y^{I})_{hor}
({\frac{q_{I}}{r^{2}}}).  \label{horizon}
\end{equation}
However, $Y_{hor}^{I}={\cal V}_{hor}^{1/3}(Y)X_{hor}^{I}$, and thus 
\begin{equation}
{\cal V}_{hor}^{2/3}(Y)={\frac{1}{3}}{\frac{(Z_{e})_{{\rm cr}}}{r^{2}}}
\end{equation}
where $Z_{e}=q_{I}X^{I}$ is the electric central charge, and $(Z_{e})_{{\rm cr}}$ is
its value at the horizon. Moreover, equation (\ref{jennie}) which defines
the moduli over space-time becomes near the horizon 
\begin{equation}
(Z_{e}X_{I})_{hor}=q_{I},
\end{equation}
which is what one obtains from the extremisation of the central charge.
Comparison of the form of our metric near the horizon with (\ref{berlin}),
implies that the two forms are related by 
\begin{equation}
r_{0}^{2}={\frac{(Z_{e})_{{\rm cr}}}{3}},\qquad {\hat{r}}={\frac{r^{2}}{%
2r_{0}}}.
\end{equation}
The proof for the enhancement of supersymmetry for the double extreme black
hole solution therefore remains unchanged for the general electric BPS
solution as the metrics near the horizon are identical.

Here we complete the analysis by solving for the Killing spinor near the
horizon. There the gravitino supersymmetry transformations become 
\begin{eqnarray}
\delta\psi_t &=&\nabla _t\epsilon=\Big(\partial_t +ir^2({\frac{3}{(Z_e )_{{\rm cr}}}})^{
\frac{3}{2}}\Gamma_1(1+i\Gamma_0)\Big)\epsilon,  \nonumber \\
\delta\psi_r&=&\nabla _r\epsilon=\Big(\partial_r+{\frac{i}{r}}\Gamma_0\Big) 
\epsilon,  \nonumber \\
\delta\psi_\theta&=&\nabla _\theta\epsilon=\Big(\partial_\theta+{\frac{i}{2}}
\Gamma_{012}\Big)\epsilon,  \nonumber \\
\delta\psi_\phi &=&\nabla _\phi\epsilon=\Big(\partial_\phi-{\frac{1}{2}}
\cos\theta\Gamma_{23}+{\frac{i}{2}} \Gamma_{013}\sin\theta\Big)\epsilon, 
\nonumber \\
\delta\psi_\psi&=&\nabla _\psi\epsilon=\Big(\partial_\psi+{\frac{1}{2}}
\sin\theta\Gamma_{24}+{\frac{i}{2}} \Gamma_{014}\cos\theta\Big)\epsilon,
\label{barbara}
\end{eqnarray}
where we have used that near the horizon $X^IF_{tr}^I=-{\frac{\textstyle 6r}{
\textstyle(Z_e )_{{\rm cr}}}}$.

The integrability conditions of the gravitino transformations impose no
conditions on the Killing spinor $\epsilon,$ which implies that one has full
supersymmetry. Setting all the variations in (\ref{barbara}) to zero can be
solved by the Killing spinor 
\begin{equation}
\epsilon (r,\theta ,\phi ,\psi )=e^{-{\frac{i}{2}}\Gamma _{012}\theta }e^{{\ 
\frac{1}{2}}\Gamma _{23}\phi }e^{-{\frac{i}{2}}\Gamma _{014}\psi }\Big({\ 
\frac{1}{2r}}-itr({\frac{3}{(Z_e )_{{\rm cr}}})}^{\frac{3}{2}}\Gamma _{1}(1+i\Gamma _{0}) 
\Big)\varepsilon_0,
\end{equation}
where $\varepsilon_0$ is a constant spinor. 
\section{BPS Magnetic Solutions}
In this section we construct general BPS magnetic string-like solutions
which break half of supersymmetry of $d=5,$ $N=2$ supergravity theory coupled to vector supermultiplets. As a general metric for the
magnetic string-solution we write 
\begin{equation}
ds^{2}=e^{2V}(-dt^{2}+dz^{2})+e^{2U}(d\vec{x})^{2},  \label{ma}
\end{equation}
where $(d\vec{x})^{2}=(dx^{1})^{2}+(dx^{2})^{2}+(dx^{3})^{2},$ and $(x^{1},
x^{2}, x^{3})$ are the transverse dimensions. The functions $U$ and $V$ are
taken to be independent of $(t,z).$ The non-vanishing F\"{u}nfbeins for the
metric in (\ref{ma}) are 
\begin{equation}
e_{{t}}^{\ 0}=e_{{z}}^{\ 1}=e^{V},\qquad\qquad e_{{m}}^{\ a}=e^{U}\delta_{m}^{\ a}.
\end{equation}
For the spin connections one obtains 
\begin{eqnarray}
\omega _{ta0} &=&\partial _{m}Ve^{V-U}\delta _{a}^{m},  \nonumber \\
\omega _{za1} &=&-\partial _{m}Ve^{V-U}\delta _{a}^{m},  \nonumber \\
\omega _{pab} &=&\partial _{n}U(\delta ^{an}\delta _{p}^{b}-\delta
^{nb}\delta _{p}^{a}).
\end{eqnarray}
First we consider the supersymmetry variation of the gauginos and 
find the conditions that must be satisfied by a supersymmetric bosonic
configuration. These conditions are obtained by setting the fermions and
their supersymmetry  variations to zero. Using (\ref{metric}), the gaugino
transformation can be rewritten in the form 
\begin{equation}
\delta \lambda _{i}=-{\frac{1}{4}}\left( G_{IJ}\partial _{i}X^{I}\Gamma
^{\mu \nu }F_{\mu \nu }^{J}-{3i}\Gamma ^{\mu }\partial _{\mu }X_{I}\partial
_{i}X^{I}\right) \epsilon.
\end{equation}
The vanishing of the gaugino transformation thus gives 
\begin{equation}
G_{IJ}\partial _{i}X^{I}\left( \Gamma ^{mn}F_{mn}^{J}+2i\partial
_{m}X^{J}\Gamma ^{m}\right) \epsilon =0,
 \label{g}
\end{equation}
where we have used the relation 
\begin{equation}
G_{IJ}\partial _{\mu }X^{J}=-{\frac{3}{2}}\partial _{\mu }X_{I},
\end{equation}
which follows from very special geometry. Assuming that the susy parameter
satisfies 
\begin{equation}
\epsilon =\Gamma _{1}\Gamma _{0}\epsilon ,  \label{bettina}
\end{equation}
a condition which implies that half of supersymmetry is broken. Using (\ref
{bettina}), it can be shown that 
\begin{equation}
\Gamma ^{mn}\epsilon =ie^{-3U}\epsilon ^{mnp}\Gamma _{p}\epsilon ,  \label{f}
\end{equation}
If one also writes $X^{I}=e^{-U}$ $H^{I}$, then (\ref{g}) together with the
relation $\ X^{I}\partial _{i}X^{I}=0,$ implies the following 
\begin{equation}
\Gamma ^{mn}F_{mn}^{I}\epsilon =-2ie^{-U}\partial _{p}H^{I}\Gamma
^{p}\epsilon.
\end{equation}
This, with (\ref{f}) fix the gauge field strengths to be 
\begin{equation}
F_{mn}^{I}=-\epsilon_{mnp}\partial_{p}H^{I}.
\end{equation}
Moreover, using the relation ${\frac{1}{6}}C_{IJK}X^{I}X^{J}X^{K}=1$, one
concludes that 
\begin{equation}
e^{3U}={\frac{1}{6}}C_{IJK}H^{I}H^{J}H^{K}.
\end{equation}
Therefore, using the relations of very special geometry, the
vanishing of the gaugino supersymmetry variation for a choice of the
supersymmetry parameter, fixes the relation between the gauge field strengths, $X^{I}$ and $U$.

The vanishing of the gravitino supersymmetry transformation fixes the metric
as well as the supersymmetry parameter itself. The vanishing of the $t$-component gives 
\begin{equation}
\delta \psi_{t}=\left( \partial_{t}+{\frac{1}{4}}\omega _{t}^{ab}\Gamma
_{ab}+{\frac{i}{8}}\Gamma _{t}^{\ mn}X_IF_{mn}^{I}\right) \epsilon=0.
\label{moon}
\end{equation}
Using our Ansatz, we have 
\begin{equation}
{\frac{1}{4}}\omega_{t}^{ab}\Gamma_{ab}\epsilon ={\frac{1}{2}}\omega
_{t}^{a0}\Gamma _{a0}\epsilon=-{\frac{1}{2}}\partial_{m}Ve^{V-U} \delta
_{m}^{a}\Gamma_{a}\Gamma_{0}\epsilon.
\end{equation}
Moreover, using the expressions for the scalar fields and gauge field
strengths we get 
\begin{equation}
{\frac{i}{8}}X_{I}\Gamma_{t}^{mn}F_{mn}^{I}={\frac{1}{4}}e^{V-U}%
\partial_{m}U\delta _{a}^{m}\Gamma_{a0} \epsilon.
\end{equation}
Therefore, from (\ref{moon}), we get the conditions 
\begin{equation}
V=-{\frac{1}{2}}U,\qquad \partial_t\epsilon=0.
\end{equation}
Similarly, the $z$ component of the gravitino supersymmetry variation gives 
\begin{equation}
\delta \psi _{z}=\left( \partial _{z}+{\frac{1}{4}}\omega _{z}^{ab}\Gamma
_{ab}+{\frac{i}{8}}X_{I}\Gamma _{z}^{mn}F_{mn}^{I}\right) \epsilon.
\end{equation}
Using $X_I\Gamma ^{mn}F_{mn}^{I}\epsilon =-2i\Gamma ^{p}\partial _{p}U\epsilon,$
$\Gamma _{z}=e_{z}^{1}$ $\Gamma_{1}$, \ $\Gamma ^{p}=e_{a}^{p}$ $\Gamma ^{a}$
, we deduce that $\partial _{z}\epsilon =0.$ Moreover, one obtains from the
vanishing of the space-components of the gravitino supersymmetry variations
\begin{equation}
(\partial _{m}+{\frac{1}{4}}\partial_{m}U)\epsilon =0.
\end{equation}
Thus the Killing spinor is given by 
\begin{equation}
\epsilon =e^{-{\frac{1}{4}}U}\epsilon_{0},
\end{equation}
where $\epsilon_{0}$ is a constant spinor satisfying $\epsilon
=\Gamma_{1}\Gamma_{0}\epsilon .$

Here we give a summary of what we have done so far. We are seeking a
background which breaks supersymmetry and we have obtained the following
configuration 
\begin{equation}
F_{mn}^{I}=-\epsilon _{mnp}\partial _{p}H^{I},\qquad
X^{I}=e^{-U}H^{I},\qquad V=-{\frac{1}{2}}U.
\end{equation}
In order to fix the scalar functions $H^{I}$, one has to solve for the
equations of motion and the Bianchi identities for the gauge fields. Doing
so, the $H^{I}$ are then fixed and are given by a set of harmonic functions 
\begin{equation}
H^{I}=g^{I}+{\frac{p^{I}}{r}},
\end{equation}
where $g^{I}$ are constants and $p^I$ are magnetic charges. As for the
electric black hole solutions, we define new coordinates $Y^{I}=e^{U}X^{I},$
in terms of which the magnetic string solution takes the following simple form 
\begin{eqnarray}
ds^{2} &=&{\cal {V}}(Y)^{-{\frac{1}{3}}}(-dt^{2}+dz^{2})+{\cal {V}}(Y)^{%
\frac{2}{3}}(dr^{2}+r^{2}d\theta ^{2}+r^2\sin ^{2}\theta d\phi ^{2}),  \nonumber \\
F_{mn}^{I} &=&-\epsilon _{mnp}\partial _{p}H^{I},  \nonumber \\
Y^{I} &=&H^{I}.
\end{eqnarray}

In the electric case, the horizon geometry is given by $AdS_{2} \times S^3$
and the black hole entropy, related to the horizon volume $S^3$, is given in
terms of the extremised electric central charge. For the magnetically
charged $d=5$ BPS black string, one similarly find that for our solution the
extremised value of the BPS tension is related to the volume of the $S^2$
where the near horizon geometry is given by $AdS_{3} \times S^2$. This can
be easily demonstrated. The magnetic central charge related to the tension
of the magnetic string states is defined by $Z_m=p^IX_I$. To find the
extremum we set
\begin{equation}
\partial_i Z_m = \partial_i(X_I)p^I = {\frac{1}{3}} C_{IJK} X^I \partial_i
(X^J) p^K =0.
\end{equation}
It follows that the critical values of $X^I$ and its dual are given by \cite
{Chou} 
\begin{equation}
X^I = {\frac{p^I}{Z_m}}, \qquad X_I = {\frac{1}{6}}{\frac{C_{IJK} p^J p^K }{%
Z^2_m}}
\end{equation}
and thus the critical value of the BPS string tension is 
\begin{equation}
(Z_m ) ^3 _{{\rm cr}} = {\frac{1}{6}}C_{IJK} p^I p^J p^K.
\end{equation}
At the near horizon $(r\rightarrow 0)$ we have 
\begin{eqnarray}
e^{3U}={\frac{1}{6}}C_{IJK}H^IH^JH^K & \rightarrow & {\frac{1}{6r^3}}
C_{IJK}p^Ip^Jp^K\equiv {\frac{(Z_m ) ^3 _{{\rm cr}} }{r^3}}  \nonumber \\
X^I=e^{-U}H^I &\rightarrow & {\frac{p^I}{(Z_m ) _{{\rm cr}} }}  \nonumber \\
X_I={\frac{1}{6}}C_{IJK}H^JH^K &\rightarrow & {\frac{1}{6}}C_{IJK}{\frac{%
p^Jp^K}{(Z_m ) ^2_{{\rm cr}} }}.
\end{eqnarray}
Therefore the values of the moduli at the horizon are those
which extremise the magnetic central charge. With these limits, the metric
and the gauge fields at the horizon could then be expressed in the form 
\begin{eqnarray}
ds^2&=&{\frac{r}{ (Z_m ) _{{\rm cr}} }}(-dt^2+dz^2)+{\frac{(Z_m ) _{{\rm cr}%
}^2}{r^2}}dr^2+(Z_m ) _{{\rm cr}}^2(d\theta ^{2}+\sin ^{2}\theta d\phi ^{2}),  \nonumber \\
F_{\theta\phi}^I &=& q^I \sin\theta.
\end{eqnarray}

We now demonstrate the enhancement of supersymmetry at the horizon for the magnetic
solutions. For the near horizon metric, the gaugino
transformation reduces to 
\begin{equation}
\delta\lambda_{i}=\Big({\frac{1}{3(Z_m ) _{{\rm cr}}^2}}\partial_{i}(q^IX_I)
\Gamma_{34}+ {\frac{3i}{4(Z_m ) _{{\rm cr}}}}r(\partial_rX_I\partial_iX^I)
\Gamma_2\Big)\epsilon  \label{amos}
\end{equation}
The above variation vanishes identically without any conditions imposed on
the supersymmetry parameter. This is basically due to the fact that the
magnetic central charge is extremised at the horizon. From the gravitino
supersymmetry transformations near the horizon we get the equations 
\begin{eqnarray}
\delta\psi_t &=&\nabla _t\epsilon=\Big(\partial_t +{\frac{r^{\frac{1}{2}}}{
4(Z_m ) _{{\rm cr}}^{\frac{3}{2}}}}(\Gamma_0-\Gamma_1)\Gamma_2\Big)\epsilon,
\nonumber \\
\delta\psi_z&=&\nabla_z\epsilon=\Big(\partial_z-{\frac{r^{\frac{1}{2}}}{%
4(Z_m ) _{{\rm cr}}^{\frac{3}{2}}}}(\Gamma_0-\Gamma_1)\Gamma_2\Big)\epsilon,
\nonumber \\
\delta\psi_r&=&\nabla _r\epsilon=\Big(\partial_r+{\frac{1}{4r}}%
\Gamma_0\Gamma_1\Big)\epsilon,  \nonumber \\
\delta\psi_\theta&=&\nabla _\theta\epsilon=\Big(\partial_\theta-{\frac{i}{2}}
\Gamma_4\Big)\epsilon,  \nonumber \\
\delta\psi_\phi &=&\nabla _\phi\epsilon=\Big(\partial_\phi+{\frac{i}{2}}%
\sin\theta\Gamma_3-{\frac{1}{2}}\cos\theta \Gamma_3\Gamma_4\Big)\epsilon,
\end{eqnarray}
The integrability conditions 
\begin{equation}
[\nabla_\mu, \nabla_\nu]\epsilon=0
\end{equation}
can be seen to be satisfied with no restrictions on the
Killing spinor $\epsilon.$ This implies that supersymmetry is fully
restored. The solution of the Killing spinor is given by \footnote{%
the dependence of the Killing spinor on $Z_m$ can be absorbed by the redefinition of coordinates, 
$r=r^{\prime} (Z_m )^3 _{{\rm cr}}$} 
\begin{equation}
\epsilon(t,z,r,\theta, \phi)= e^{{\frac{i}{2}}\Gamma_4\theta} e^{{\frac{1}{2}
}\Gamma_3\Gamma_4\phi}\Big((1+\Gamma_0\Gamma_1)r^{\frac{-1}{4}}-{\frac{(t-z)
}{2(Z_m ) _{{\rm cr}}^{\frac{3}{2}}}}r^{\frac{1}{4}}(\Gamma_0-\Gamma_1)
\Gamma_2\Big)\kappa_0.
\end{equation}
where $\kappa_0$ is a constant spinor.
The structure of the horizon is $AdS_3\times S^2$ with curvatures 
\begin{eqnarray}
R_{ab}&=&-{\frac{1}{4(Z_m ) _{{\rm cr}}^2}}\eta'_{ab}, \qquad \qquad
\eta'_{ab}=(-1, 1, 1),\qquad \qquad a,b=0,1,2  \nonumber \\
R_{\alpha\beta}&=&{\frac{1}{2(Z_m ) _{{\rm cr}}^2}}\delta_{\alpha\beta},
\qquad \qquad \ \alpha,\beta=3,4.
\end{eqnarray}

In conclusion, with the help of the relations of very special geometry, BPS
electric and magnetic solutions of M-theory on a Calabi-Yau manifold were
obtained. Our analysis applies to all $N=2$ five dimensional
supergravity and not only to models with higher-dimensional origin, $i. e, $ those obtained from compactifying 
11-dimensional supergravity on a Calabi-Yau manifold. The
solutions depend on the prepotential which defines the theory. We note that
in the electric solutions, the rescaled four-cycles $Y_I$ are given in terms
of harmonic functions and thus in order to find the explicit space-time
solution one has to solve for the two-cycles $X^I$. In the magnetic case,
however, it is the rescaled two-cycle which is given in terms of a harmonic
function and thus the explicit space-time solution is known for all
prepotentials. We have demonstrated that supersymmetry in enhanced at
the near horizon of the general electric and magnetic solutions. The
solutions for the Killing spinors of the near horizon geometries, $
AdS_2\times S^3$ and $AdS_3\times S^2$ are given. 
\vfill\eject

\end{document}